\documentclass[11pt]{article}
\usepackage{moriond}
\usepackage{amsmath}
\usepackage{array}
\usepackage{color}
\usepackage{graphicx}
\usepackage{subfigure}
\usepackage{multirow}
\usepackage{wrapfig}

\graphicspath{{.}}
\DeclareGraphicsExtensions{ .eps , .eps.gz ,
                            .ps  , .ps.gz }
\newcommand{\gm}{$(g-2)$}

\newcommand\Journal[4]{{#1} {\bf #2}, #3 (#4)}
\newcommand\RMP{Rev. Mod. Phys.}

\newcommand\PR{Phys. Rev.}

\newcommand\EPC{{Eur. Phys. J.} C}
\newcommand\NIMA{Nucl. Inst. and Meth. A}
\newcommand\NPS{Nucl. Phys. B (Proc. Suppl.)}
\newcommand\NPB{ Nucl. Phys. B}

\newcommand\PRL{Phys. Rev. Lett.}
\newcommand\PRD{Phys. Rev. D}

\newcommand\ea{{\em et al.}}

\renewcommand{\vec}{\boldsymbol}

\begin{document}
\vspace*{4cm}

\title{A NEW PRECISE MEASUREMENT OF THE\\ MUON ANOMALOUS MAGNETIC MOMENT}

\author{GERCO ONDERWATER\\ \vspace{1mm}{\small for the Muon g-2
Collaboration~\cite{collab}}}

\address{University of Illinois at Urbana-Champaign, \\
         1110 West Green Street, Urbana, IL 61801, USA
         \begin{figure}[h]
         \begin{center}
         \includegraphics*[height=4cm] {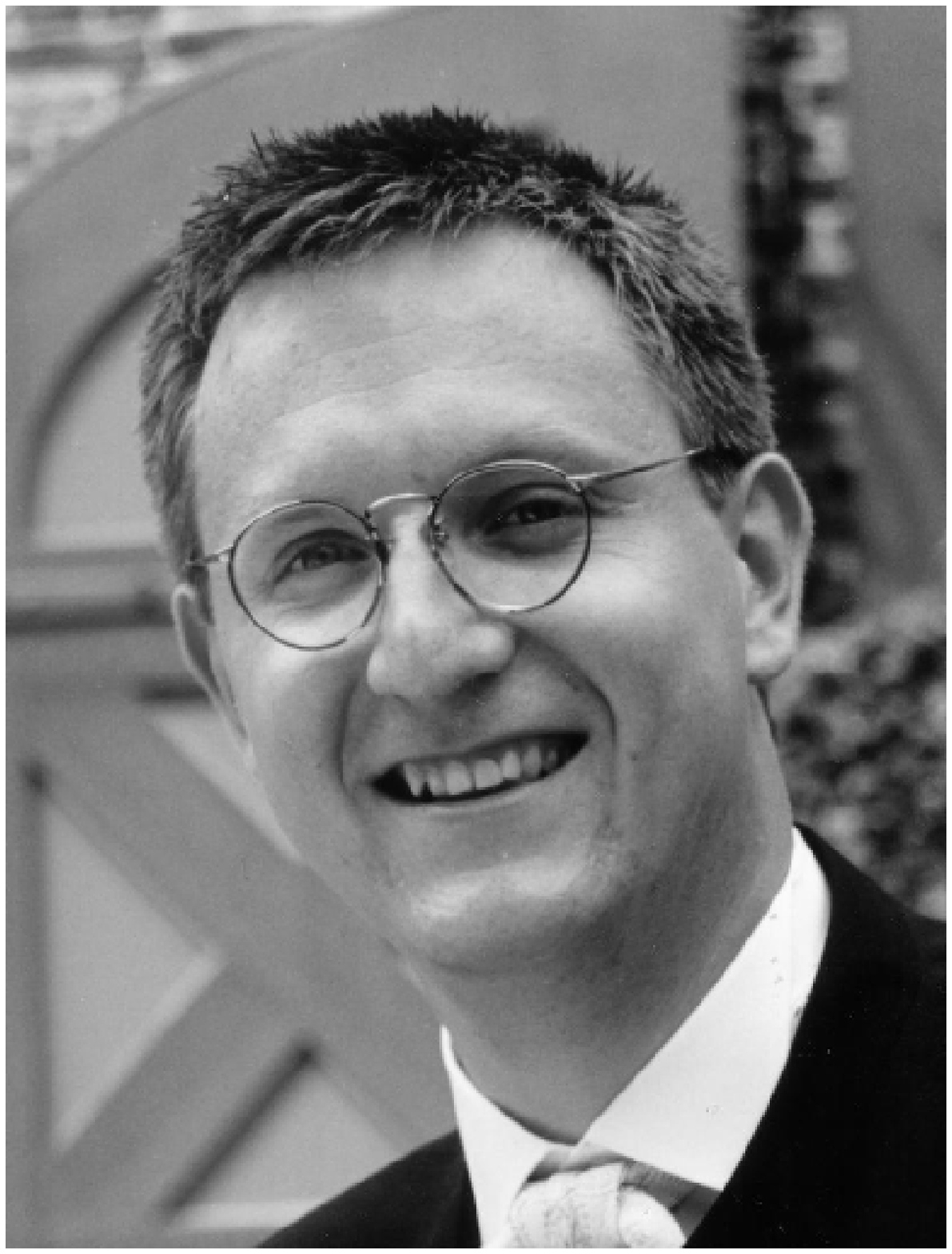}
         \end{center}
         \end{figure}
         \vspace{1em}
}

\maketitle\abstracts{
  The muon \gm\ experiment at BNL has completed four data runs for the
  positive muon, beginning in 1997 and has just finished its first run
  for the negative muon.  Recently the analysis of a 1.3~ppm result
  based on the 1999 run was completed and published ($a_{\mu^+}~=
  11\,659\,202(14)(6)\times10^{-10}$).  The result shows a tantalizing
  discrepancy with the Standard Model prediction of 2.6 times the
  combined experimental and theoretical uncertainty.  The muon \gm\
  experiment at BNL is ultimately aimed at measuring the anomalous
  magnetic moment of the muon with a relative uncertainty of 0.35~ppm. 
}

\section{Introduction}

The quest for phenomena that cannot be explained by the Standard Model
(SM) of fundamental interactions is a focal point of present-day
particle physics.  Although the SM is consistent with all available
experimental data, there are increasingly strong theoretical reasons
and also circumstantial evidence that point to the existence of new
physics below the TeV scale~\cite{herczeg}.  Precision experiments at
low energy can probe for deviations from the SM predictions and
provide complementary constraints to those obtained in high-energy
collision experiments.  The uncertainty principle mandates that even
heavy particles --- which may not yet have been discovered --- make
small contributions to low-energy observables, such as the magnetic
moment of the muon.

In this paper, we report on the latest result~\cite{brown2001} of
experiment E821, which is ongoing at Brookhaven National Laboratory
and aims to measure the muon magnetic moment anomaly $a_\mu$ with a
relative uncertainty of 0.35 parts-per-million (ppm).  The experiment
has been taking data since 1997 with intermediate
results~\cite{carey,brown}
\begin{eqnarray}
  a_{\mu^+}(1997) &=& 1~165~925(15)\times 10^{-9}~(13~\text{ppm})\\
  a_{\mu^+}(1998) &=& 1~165~920(6)\times 10^{-9}~(4~\text{ppm})
\end{eqnarray}
in good agreement with earlier measurements~\cite{cern3} conducted at
CERN
\begin{eqnarray}
  a_{\mu^+} &=& 1~165~911(11)\times 10^{-9}~(10~\text{ppm})\\
  a_{\mu^-} &=& 1~165~937(12)\times 10^{-9}~(11~\text{ppm})
\end{eqnarray}
The combination of these results is consistent with the Standard Model
prediction within the combined experimental and theoretical
uncertainty.

\section{Theory}

The magnetic moment of a particle is related to its spin,
\begin{equation}
	\vec \mu = g \left(\frac{e}{2mc}\right) \vec S.
\end{equation}
According to Dirac's theory, the proportionality constant $g$, or
gyromagnetic ratio, is equal to 2 for point-like, {\em i.e.}
structureless, particles.  For charged baryons $g$ differs
substantially from 2, reflecting their internal structure.  The
electron and muon are special in that they appear to be point-like. 
Nevertheless, it was discovered that $g_e$ is slightly larger than 2
for electrons~\cite{kusch}.  The deviation, referred to as the
magnetic moment anomaly $a$, is defined through the
relation~\cite{pdg}
\begin{equation} 
  \mu = 2 (1 + a) {e \hbar \over 2 m} \qquad {\rm where}
  \qquad a = {(g-2)\over 2}.
  \label{eq:mom}
\end{equation} 
The anomaly is the radiative correction to the magnetic moment, and
was first calculated in 1948 by Schwinger~\cite{schwinger}, predicting
that, to first order, $a = \alpha/2\pi \simeq 1/800$.  During more
than 40 years, increasingly refined experiments and equally
sophisticated calculations have made the anomalous magnetic moment of
the {\em electron} one of the best understood physics phenomena.  The
electron anomaly is now measured with an experimental uncertainty of a
few parts-per-billion~\cite{ge} and is well described by QED
calculations~\cite{kinhughes}.

While the $g$-factor of the electron has provided a testing ground for
QED, the anomalous magnetic moment of the muon has provided an even
richer source of information, because the magnitude of radiative
corrections involving heavy virtual particles scales with the square
of the lepton mass and $\left(m_\mu/m_e\right)^2 \simeq 40,000$.  The
contributions of virtual muons and hadrons were already demonstrated
in the experiments at CERN, which have a combined uncertainty of
7.3~ppm for $a_\mu$.  This uncertainty was not sufficient to observe
the contribution of the carriers of the weak interaction, the $W^\pm$
and $Z^0$ bosons.  The goal of experiment E821 is an overall
uncertainty in $a_\mu$ of 0.35~ppm, which will allow the observation
of the weak contributions, as well as a search for physics beyond the
SM.

The SM prediction for the anomaly $a_\mu$ is traditionally calculated
as the sum of three parts~\cite{czar},
\begin{equation}
  a_{\mu}{(\text{SM})} = a_{\mu}(\text{QED}) + a_{\mu}(\text{hadronic}) +
      a_{\mu}(\text{weak}).
\end{equation}
The presently most accurate predictions for each of these
contributions given in table~\ref{table1} were taken from
Ref.~\cite{ourtheory} and references therein.  The uncertainty is
dominated by the hadronic contribution, which cannot be calculated
from first principles and is (mostly) determined from
data~\cite{hocker}.

\begin{table}[ht]
 \caption{Standard Model predictions for $a_{\mu}$~{\protect\cite{ourtheory}}.\label{table1}}
  \vspace{0.4cm}
  \begin{center}
    \begin{tabular}{|l|r|r|}
      \hline
      Variable & Value ($\times 10^{10}$) & relative contribution (ppm)\\
      \hline
      $a_{\mu} (\text{SM})$ & $116\,591\,59.6(6.7)$      & $\pm 0.57$ \\
      \hline
      $a_{\mu}(\text{QED})$  & $116\,584\,70.6\,(0.3)$ & $10^6  \pm 0.03$ \\
      $a_{\mu}(\text{hadr})$ &          $673.9\,(6.7)$   & $57.79 \pm 0.57$ \\
      $a_{\mu}(\text{weak})$ &           $15.1\,(0.4)$   &  $1.30 \pm 0.03$ \\
      \hline
    \end{tabular}
  \end{center}
\end{table}
Any contribution from ``new'' physics will be reflected in a measured
value which deviates from this prediction.  There is increasing
theoretical interest in possible non-SM contributions to the muon \gm\
value.  Under suitable conditions, the final experimental and
theoretical uncertainty will allow to set more stringent limits on
possible SM extensions than obtained so far.

\section{Principle of the Experiment}

The experimental determination of the anomalous magnetic moment is
derived from the rate at which the muon's spin precesses in a magnetic
field.  When in flight, the spin precession frequency {\em relative}
to the orbital (cyclotron) frequency is proportional to the product of
the {\em anomalous} part of the magnetic moment and the strength of
the magnetic field.  The relative frequency is thus a factor 800 more
sensitive to $a_\mu$ than the spin precession frequency itself.

Ideally, polarized muons are stored in a uniform magnetic field $\vec
B$ which is perpendicular to the spin direction.  The spatial
uniformity of the magnetic field is essential in avoiding the need for
the precise measurement of the muon distribution.  Thus the use of a
magnetic quadrupole for vertical confinement, which a uniform magnetic
field cannot provide, is prohibited.  Vertical containment is
therefore established with an electric quadrupole field $\vec E$.  The
relative spin precession frequency, $\omega_a$, in the presence of
both magnetic and electric fields is given by~\cite{cern3}
\begin{equation}
   \vec \omega_a = - {e \over m }\left[ a_{\mu} \vec B -
   \left( a_{\mu}- {1 \over \gamma^2 - 1}\right)
   \vec \beta  \times \vec E \right].
   \label{eq:omega}
\end{equation}
For muons with the ``magic'' momentum $\gamma_m = \sqrt{1+1/a_\mu}
\simeq 29.3$ ($p_{\mu}$ = 3.094 GeV/$c$), the dependence of $\omega_a$
on the electric field vanishes.  The need to measure the electric
field precisely is thus avoided and precise measurements of only $B$ and
$\omega_a$ are sufficient to determine $a_\mu$.  It is important to
note that these quantities need to be known accurately {\em averaged
over the ensemble of detected events}, and not for individual muons.

The concept of experiment E821 is similar to that of the last
experiment at CERN, which is described in detail in Ref.~\cite{cern3}. 
Major improvements that were made compared to the CERN experiments
are\\
\parbox{0.5\textwidth}
{
\begin{itemize}
  \item Higher beam intensity;
  \item Direct muon injection;
  \item A DC superconducting inflector;
  \item The quadrupole design.
  \item Circular storage aperture;
\end{itemize}
}
\parbox{0.5\textwidth}
{
\begin{itemize}
  \item {\em In vacuo} field measurement;
  \item Waveform Digitizers;
  \item Super-ferric magnet;
  \item Higher resolution detectors;
  \item Scalloped vacuum chambers.
\end{itemize}
}
In the following sections, some of the details of the experiment are
described.

\section{Creating and Storing Polarized Muons}

The Alternating Gradient Synchrotron (AGS) accelerates up to 12
bunches of protons to an energy of 24.3~GeV in cycles of 2~--~3
seconds.  Every 33~ms one 25~ns wide (FWHM) bunch is extracted and
directed to a production target to create pions.  Only pions with a
momentum 1.7\% above the magic momentum of 3.094~GeV/$c$ are
transported through an approximately 100~m long beamline in which they
will decay, $\pi^\pm \rightarrow \mu^\pm + \nu_\mu$.  In the pion's
center-of-mass frame, the muon's spin is aligned with its momentum
because of the pion's spin ($S$=0) and the uniquely defined handedness
of the neutrino.  Consequently, in the laboratory frame the highest
energy muons are nearly 100\% polarized.  At the end of the beamline,
muons at the magic momentum are selected by a pair of dipole magnets
and collimators.

The resulting bunch of about 50,000 muons is injected into a 14.2~m
diameter super-ferric storage ring~\cite{danby} through a
superconducting inflector channel.  The purpose of the inflector is
twofold.  First, its superconducting shield cancels the magnetic field
of the main magnet while the beam traverses its yoke and approaches
the storage orbit as close as possible.  Second, a set of carefully
wound coils~\cite{krienen} counter-balances the effect the field-free
region has on the magnetic field in the storage area.

The main field of 1.45~T is excited by three helium-cooled
superconducting coils and contained in an C-shaped iron yoke.  The
uniformity of the field is controlled by carefully shaped and aligned
pole piece segments and further optimized by inserting wedges into the
air gap between the yoke and the poles.  Minimization of the azimuthal
variation in the field was accomplished by inserting thin
ferromagnetic plates between the poles and the vacuum chamber.  The
transverse field variations are reduced by 2$\times$120 wires running
azimuthally around the ring on the top and bottom pole surfaces and by
local current loops.

The vacuum chamber inside the magnet maintains a pressure of
$10^{-6}$~---~$10^{-7}$~Torr and contains several circular 9~cm
diameter collimators to define the storage volume.  A circular
aperture reduces the higher multipole moments of the stored beam
profile.  Since the average magnetic field is determined by the
convolution of the magnetic field and beam distribution, this reduces
the need to measure the latter to high precision.  The chambers are
scalloped at the inner radius to reduce pre-showering and thus help to
improve the resolution with which the decay electron energy can be
measured.

Direct muon injection into the storage ring is accomplished by
deflecting the muon beam by 10~mrad at a quarter of a betatron
wavelength from the inflector.  This deflection is generated by a set
of three magnetic kicker modules.  Each module comprises a current
loop, which guarantees a negligible effect on the main magnetic field
after 20~$\mu$s.  The width of the pulse generated by the kickers is
about 150~ns FWHM, which matches the revolution time of the ring.

In order to accommodate the kicker and inflector, the electric
quadrupole field is created by four symmetrically positioned
39$^\circ$ (in azimuth) sections of two horizontal and two vertical
plates each that yield a nearly pure quadrupole potential. 
Discharging caused by the buildup of trapped electrons originating
from ionizing the residual gas in the vacuum chamber is prevented by
operating the quadrupoles in pulsed mode.  For up to 1.4~ms, a
20~---~25~kV voltage is applied at each plate (opposite polarities for
horizontal and vertical plates).  The setting and stability of the
electric field is measured to 0.2\%, which is sufficient to stay far
away from spin and beam resonances.  At this voltage, weak focusing
is achieved with a field index of $n\simeq 0.135$.

During the first 15~$\mu$s, the tails of the beam distribution are
scraped off by lowering the voltage of the inner and bottom plates. 
This will move the beam closer to the collimators, so that the edges
of the distribution, which are likely to be lost later and introduce a
systematic error, are removed.

\section{Magnetic Field}

The basic method of field measurement is the well-known proton NMR
technique~\cite{prigl}, which permits the absolute measurement of a
local magnetic field with a precision of 0.1~ppm.  The proton sample
of the probes (either water or Vaseline) is ``excited'' with a
10~$\mu$s RF pulse.  The resulting free induction decay signal, which
is proportional to the magnetic field strength, is measured with
respect to a 61.74~MHz reference by counting the number of
zero-crossings at the difference frequency of 10~---~50~kHz.  A
LORAN-C receiver serves as a frequency standard and has a long-time
stability of $10^{-12}$ and a short-time stability of better than
$10^{-10}$.

Four different kinds of probes are used:
\begin{enumerate}
\item 17 probes are mounted on a movable trolley to measure the field
      in the storage area while under vacuum;
\item 360 fixed probes are mounted along the ring for field monitoring
      purposes and (a subset) for field regulation;
\item a plunging probe serves as a secondary standard for calibrating
      the trolley;
\item a standard probe is used for absolute calibration.
\end{enumerate}

The trolley probes are mounted concentrically to cover most of the
storage aperture.  By driving the trolley through the vacuum chamber,
the field is mapped at any azimuthal position.  The transverse
position of the trolley is derived from the alignment of the rails on
which it rides and the azimuthal location is derived from the pulling
cable and the observation of the field disturbance with the fixed
probes when the trolley passes.

The fixed probes monitor time-dependent variations of the relative
field strength between the trolley measurements.  The final dependence
was obtained by taking a weighted average of about 135 probes.

Relative calibration is obtained by measuring the field at the
location of each of the trolley probes with a plunging probe.  The
plunging probe itself as well as some of the trolley probes are
absolutely calibrated with respect to the standard probe~\cite{wei}. 
This is only done at the beginning and end of a data taking period
since it cannot be done in vacuum.  The same calibration probe was
also used in the muonium hyperfine experiment for determining
$\lambda$ (see below), which eliminates (part of) the systematic error
related to absolute calibration.
\begin{figure}[ht]
  \begin{center}
  \subfigure[field map]{\label{fig:multipole}
  \includegraphics*[width=0.4\textwidth] {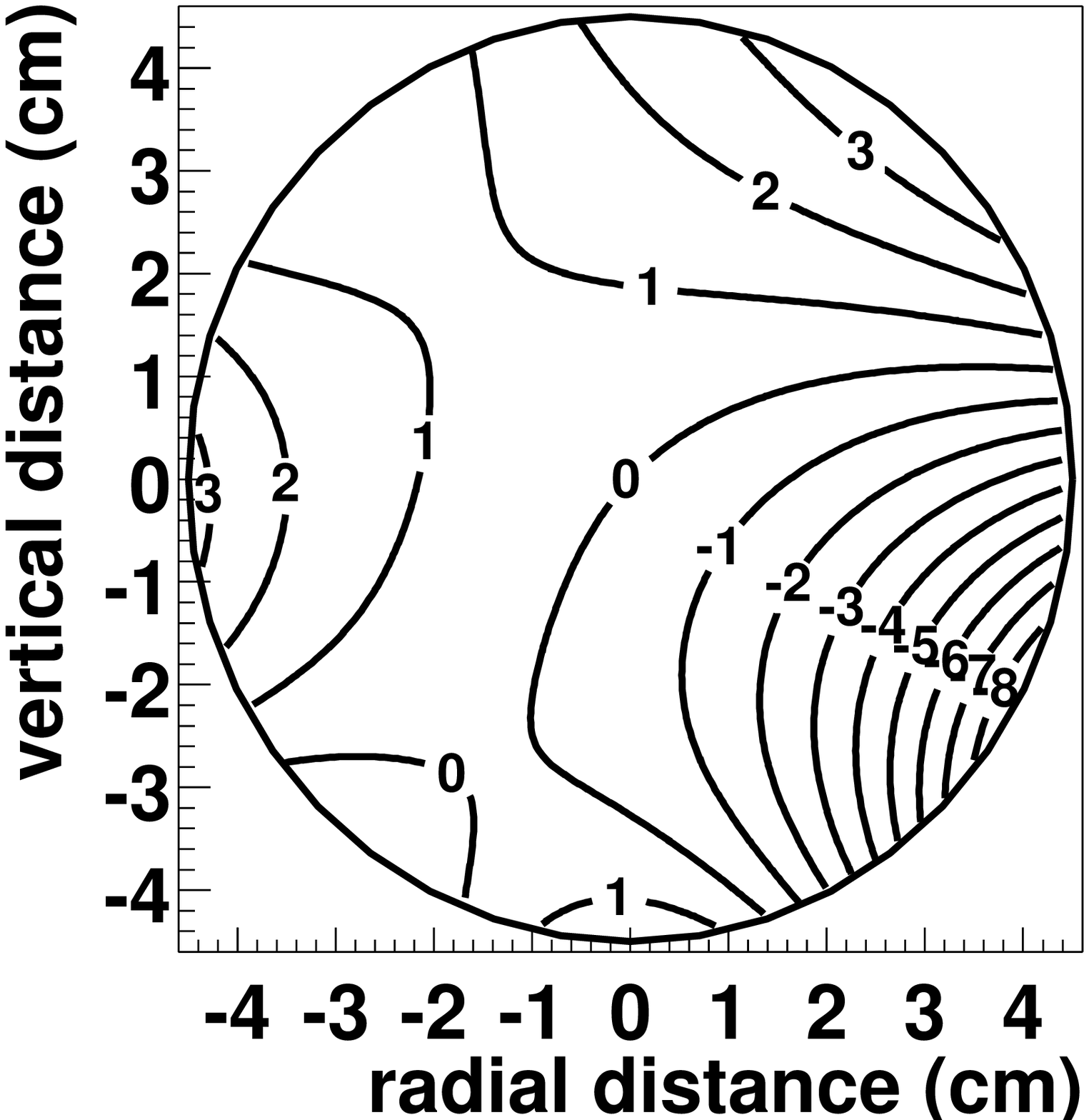}}
  \quad
  \subfigure[azimuthal field variation]{\label{fig:multipole2}
    \includegraphics[width=0.4\textwidth]{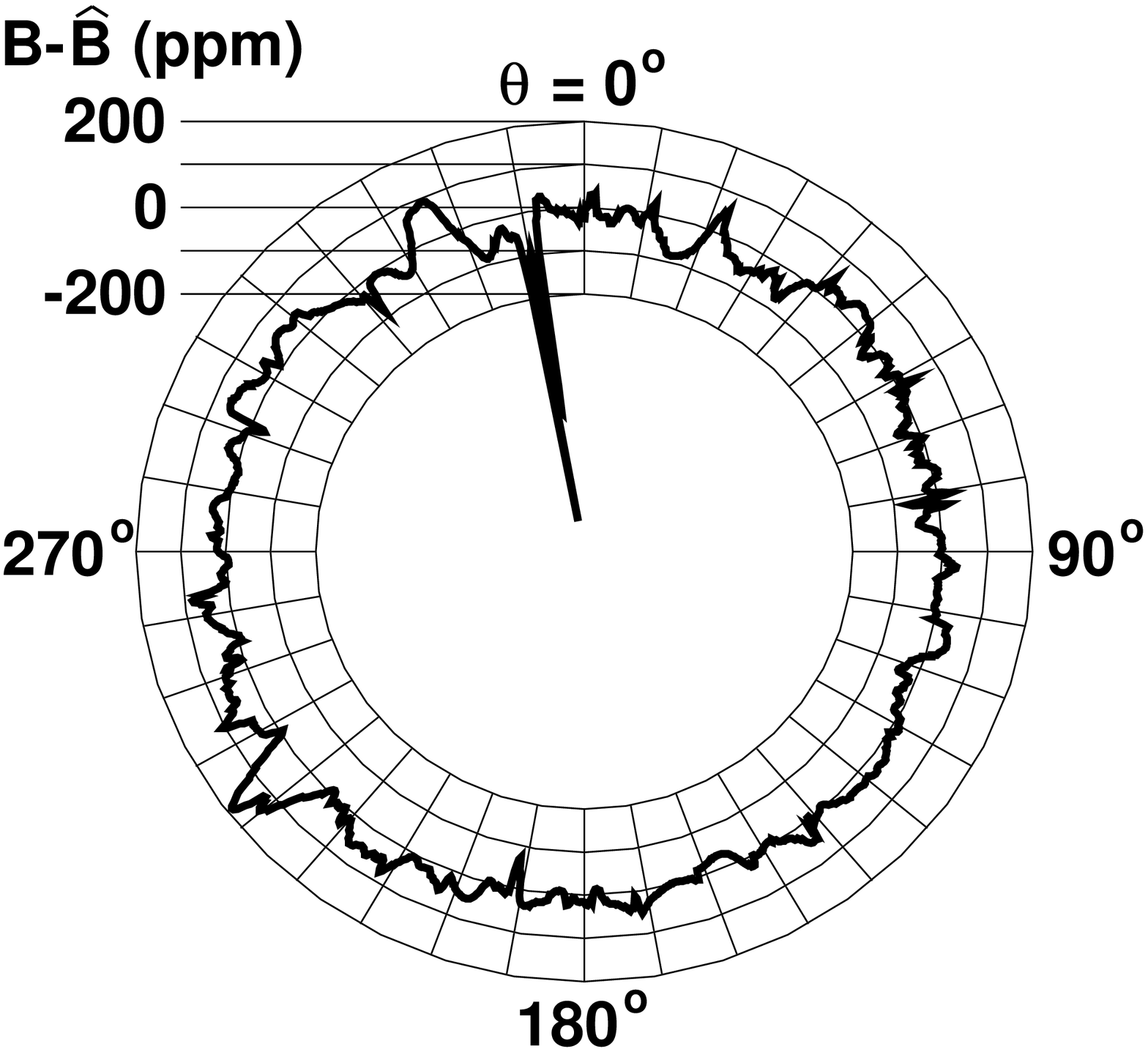}}
  \end{center}
  \caption{Typical results for one out of seventeen trolley
           measurements.  (a) Field profile averaged over azimuth and
           interpolated using a multipole expansion.  One ppm contours
           indicate variations with respect to the central value.  The
           circle covers the storage aperture.  (b) Field variations
           $dB$ measured with the trolley center probe versus azimuth
           angle $\theta$.  The inflector ``spike'' at 350$^\circ$ is
           absent in later runs (see text). 
           \label{fig:fieldmeasurement}}
\end{figure}

Figure~\ref{fig:fieldmeasurement} shows the result of a typical field
measurement.  The region near the inflector ($\theta\approx 350^\circ$
; see fig.~\ref{fig:multipole2}) shows the fringe field caused by
incomplete field compensation by the inflector, which was replaced for
future data runs.  This fringe field is responsible for the relatively
large non-uniformity in the lower right of fig.~\ref{fig:multipole}
and constitutes one of the larger systematic errors, together with the
calibration of the trolley probes and the interpolation using the
fixed probes.  Two largely independent analyses, which agreed to
within 0.03~ppm, yielded the final result $\omega_p/2\pi$ =
61,791,256~$\pm$~25~Hz (0.4~ppm).  This is the field at the measured
average beam center instead of averaged over its distribution.  The
resulting uncertainty, as well as other systematic uncertainties, are
specified in table~\ref{tab:omegapsyserr}.
\begin{table}[ht]
\begin{center}
\caption{Systematic uncertainties for the magnetic field
         analysis.}\label{tab:omegapsyserr}
\vspace{0.2cm}
  \begin{tabular}{|l|c|}
  \hline
  Source of Uncertainties & Size [ppm] \\
  \hline
  Calibration of trolley probes & 0.20 \\
  Inflector fringe field & 0.20 \\
  Uncertainty from the muon distribution & 0.12 \\
  Interpolation with fixed probes & 0.15 \\
  Trolley measurements of $B_o$ & 0.10 \\
  Absolute calibration of standard probe & 0.05 \\
  Others $\dagger$ & 0.15 \\
  \hline
  Total Uncertainty & 0.4 \\
  \hline
  \end{tabular}
\end{center}
\vspace*{3pt}
\centerline{\parbox{9cm}{\footnotesize 
$\dagger$ higher multipoles, trolley temperature and its power supply
voltage response, and eddy currents from the kicker.}}
\end{table}

\section{The Spin Precession Frequency}

The muon is an unstable particle with an intrinsic lifetime of about
2.2~$\mu$s, so the stored muons will eventually decay through the weak
interaction, $\mu^+ \rightarrow e^+ \nu_e \bar{\nu}_\mu$.  The angular
distribution of decay electrons in the center-of-mass (CM) system of
the muon is spin dependent,
\begin{equation}
  \frac{d\mathcal{N}}{d\epsilon\,d\Omega } = \mathcal{N}(\epsilon)\left[1 +
      \mathcal{A}(\epsilon)\,\cos\theta \right].
\end{equation}
Here, $\theta$ is the angle between the muon spin and the electron
momentum.  Both the phase space factor $\mathcal{N}(\epsilon)$ and the
parity violating asymmetry $\mathcal{A}(\epsilon)$ are energy
dependent.

When boosted to the laboratory (LAB) system, a strong correlation
between the emission angle in the CM and the energy in the LAB arises. 
The highest LAB energy electrons are emitted along the direction of
the muon momentum.  Consequently the rate of these electrons is
modulated depending on the orientation of the (rotating) spin with
respect to the muon momentum.  Moreover, the modulation is purely
sinusoidal, leading to an observed rate
\begin{equation}
  N(E,t) = N(E) e^{-t/\gamma\tau} \left[ 1+ A(E)\cos(\omega_a t +\phi(E))  \right].
  \label{eq:wigglefunc}
\end{equation}
Here, $N(E)$ and $A(E)$ are the LAB equivalent of
$\mathcal{N}(\epsilon)$ and $\mathcal{A}(\epsilon)$, $\omega_a$ is the
relative spin precession frequency and $\gamma\tau$ is the dilated
lifetime of 64.4~$\mu$s.  The energy dependent phase $\phi(E)$ arises
from the different flight path lengths of the decay electrons.

The energy and arrival time of the decay electrons were measured using
24 lead/scintillating-fiber calorimeters~\cite{sedykh}, placed at the
inner radius of the ring.  The signals of the four photo-multiplier
tubes reading out each calorimeter block were digitized by custom
400~MHz waveform digitizers (WFD), which use the same LORAN-C
frequency receiver as used for the magnetic field measurement.

In the first stage of the analysis, the arrival time and energy of the
decay electrons were reconstructed from the traces recorded by the
WFDs using two independent reconstruction programs.  Special effort
was put in understanding the effect of overlapping events and of
pulses with an amplitude below the software threshold.

In the next stage of the analysis, several tests were made to ensure
the quality of the data runs, AGS cycles and individual fills.  Tests
were made for sparking in the quadrupoles, failed extraction of the
AGS, misfiring of the kicker and readout and reconstruction problems. 

\begin{wrapfigure}{r}{0.5\textwidth}
  \includegraphics*[width=0.5\textwidth]{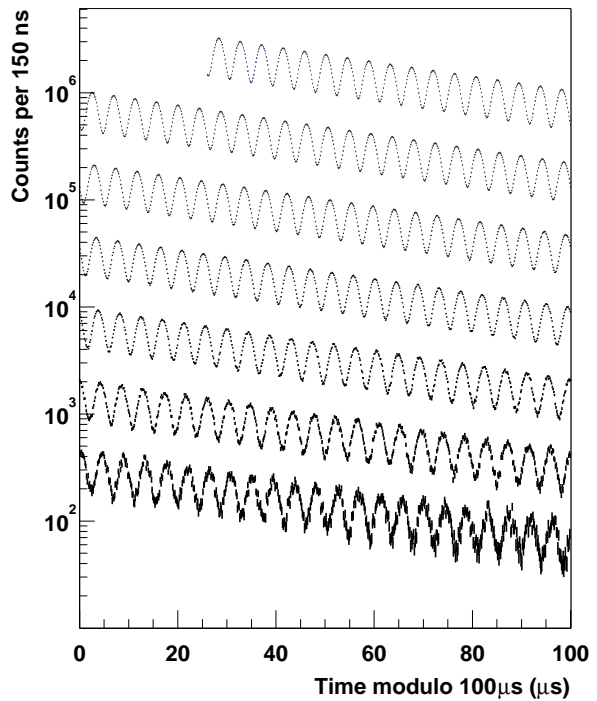}
  \caption{Observed electron rate in successive 100~$\mu$s
           periods.\label{fig:wiggleplot}}
\end{wrapfigure}
The statistical uncertainty with which $\omega_a$ can be determined
from the electron rate modulation is optimal for an energy threshold
of 1.8~GeV.  For reasons discussed below, a slightly higher threshold
of 2~GeV was used to build histograms of the number of observed
electrons as a function of time.  The resulting histogram containing
the complete 1999 data set of about 900 million events is shown in
fig.~\ref{fig:wiggleplot}.  During the first 25~$\mu$s some of the
detectors are still gated off because of the injection flash caused by
beam pions.

Following injection, the rate is modulated at the cyclotron frequency
because of the narrow width of the injected muon bunch, which only
debunches very slowly in several hundreds of turns.  This modulation
is eliminated by introducing an artificial resolution and by binning
the data in cyclotron periods.  In dedicated studies, the revolution
time and debunching characteristics were used to extract the central
momentum (and thus beam center) and momentum distribution.

Fitting eq.~(\ref{eq:wigglefunc}) to the data does not yield an
acceptable result.  Deviations of the data from this ideal function
form were identified to be caused by pileup, beam dynamics and gain
instability.  These deviations mostly cause an increase in the fitting
$\chi^2$, and affect the fitted value of $\omega_a$ only modestly. 
Fits are based on minimizing $\chi^2$ because log-likelihood
($\log(\mathcal{L})$) maximization doesn't provide a measure of the
goodness-of-fit.  We define $\chi^2 = \sum_{bins}
(N_i-f(t_i))^2/f(t_i)$, with $N$ the bin content of the histogram and
$f$ the fitted function at the center of the bin.  Only bins with 40
or more counts were used in which case $\log(\mathcal{L})$
maximization and $\chi^2$ minimization yield equivalent results.

Four complete and independent analyses were made.  In three of the
four analyses, the data were fitted using a multiparameter function,
based on eq.~(\ref{eq:wigglefunc}) with extensions to account for the
deviations.  In the fourth analysis, the so-called ratio method, the
data was manipulated, so that it could be fitted with a function
containing only three parameters.  The sensitivity of the latter
method to beam dynamics and gain instability is strongly suppressed
compared to the other methods.

In three of the four analyses, pileup was corrected for in a
statistical way.  The linear dependence of the amount of pileup on the
deadtime allows to extrapolate to zero deadtime from two data sets
with different deadtimes.  This is accomplished by extending the
intrinsic pulse-finder deadtime in software.  In the other analysis,
the contribution of pileup was added to the function and fitted.

Two beam dynamics related distortions were identified.  First, muons
are being lost other than by decay.  The direct influence on the spin
precession frequency was negligible, but a change in the exponential
decay was observed.  In an independent study, based on a set of
scintillator strips in front of the calorimeters, it was found that
the muon loss rate relative to the decay rate could be approximated as
exponential in combination with a constant (1\%).

The second beam dynamics related effect was caused by coherent
betatron oscillations (CBO), which occurred because the aperture of the
inflector is smaller than the storage aperture, and which were
enhanced by an incomplete injection kick.  The radial movement and
breathing of the beam and the varying radial acceptance of the
calorimeters led to an easily identified modulation of the electron
rate at a frequency of 470~kHz, which is the difference of the
cyclotron and horizontal betatron oscillation frequencies.

Independent studies of the beam dynamics were made using a set of
scintillating fiber harps, which could be inserted into the storage
area.  Both horizontal and vertical oscillations and breathing were
observed, as well as a small contamination with protons.  Similar
studies were made using four straw chambers, that allowed to trace the
electron back to the point at which the muon decayed and thus provided
information on the beam distribution and its motions.  All three
multiparameter fitting method incorporated a rate modulation at the
CBO frequency with a gaussian envelope with a time constant of about
100~$\mu$s.  It is noteworthy to mention that the CBO phase changes by
2$\pi$ going around the ring, so that when all detectors are added
together, the rate modulation is strongly reduced.

The gain and timing-stability of the detector system were monitored
using a fiber-distributed laser system and a set of two photodiodes
that served as references.  The gain variation of less than 0.1\% led
to a change in $\omega_a$ of less than 0.04~ppm and is further reduced
when detectors are added.  Nevertheless, in two of the three
multi-parameter analyses, a correction was applied for gain
variations.  In one, a time dependent energy threshold was applied,
whereas in the other the effect of gain variations was accounted for
in the fitting function.  The time-instability attributed to less than
0.1~ppm.
\begin{table}[ht]
  \begin{center}
  \caption{Comparison of the four analyses. \label{tab:wacomparison}}
  \begin{tabular}{|cccccc|}
  \hline
  Institute & Production$^\star$ & Fit Start ($\mu$s) & \# Fit Par.$^\dagger$ & $\chi^2/NDF$ & $R^\ddagger$ (ppm)\\
  \hline
  BU & PAW & 32 & 13 & $1.012\pm0.023$ & $119.54\pm1.24$ \\
  BNL & PAW & 32 & 10 & $1.005\pm0.023$ & $119.38\pm1.24$ \\
  UIUC & ROOT & 25--56 & 9 & $1.016\pm0.005$ & $119.60\pm1.23$ \\
  UM & ROOT & 34 & 3 & $0.986\pm0.025$ & $119.67\pm1.28$ \\
  \hline
  \multicolumn{6}{l}{\footnotesize $^\star$ the PAW and ROOT analysis package formed the basis
for the complete analysis}\\
  \multicolumn{6}{l}{
  \parbox{15cm}{\footnotesize $^\dagger$
  13 fit par: CBO, lost muons, pileup fitted; 
  10 fit par: CBO, lost muons, pileup corrected; \\
  9 fit par: CBO, lost muons fixed, pileup corrected; 
  3 fit par: ratio of shifted time spectra
  }}\\
  \multicolumn{6}{l}{\footnotesize $^\ddagger$ $\omega_a = 2\pi\times 229.1( 1 - R \times 10^{-6})$~kHz}
  \end{tabular}
  \end{center}
\end{table}

In table~\ref{tab:wacomparison}, the four analyses and their results
are compared.  The analyses underwent several internal consistency
checks and were all found to be self-consistent once effects that were
unaccounted for in the fitting function were considered.  The
differences between the four results are well within 0.5~ppm, which is
the amount by which they were expected to differ based on the data
overlap.

All four $\omega_a$ analyses were combined, while accounting for the
data overlap, to give $\omega_a/2\pi$~= 229072.8$\pm$0.3~Hz~(1.3~ppm). 
This includes a correction of +0.81$\pm$0.08~ppm for the effects of
the electric field at the muon beam center, and of vertical betatron
oscillations tilting the instantaneous spin precession vector.  The
final result is dominated by the statistical uncertainty.  The main
systematic uncertainties, which were estimated from independent
studies and simulations, are listed in table~\ref{tab:omegaasyserr}.
\begin{table}[ht]
  \begin{center}
    \caption{Systematic uncertainties in $\omega_a$. 
             \label{tab:omegaasyserr}}
    \begin{tabular}{|l|c|}
    \hline
    Source of uncertainty & Size [ppm] \\
    \hline
    Pileup                        & 0.13~ppm \\
    AGS background                & 0.10~ppm \\
    Lost muons                    & 0.10~ppm \\
    Timing Shifts                 & 0.10~ppm \\
    E field and vertical CBO      & 0.08~ppm \\
    Binning and fitting procedure & 0.07~ppm \\
    Coherent betatron oscillations & 0.05~ppm \\
    Beam debunching               & 0.04~ppm \\
    Gain instability              & 0.02~ppm \\
    \hline
    Total Uncertainty             & 0.3~ppm \\
    \hline
    \end{tabular}
  \end{center}
\end{table}

\section{Results and Conclusion}

\begin{figure}[h]
  \centerline{\includegraphics*[width=0.55\textwidth]{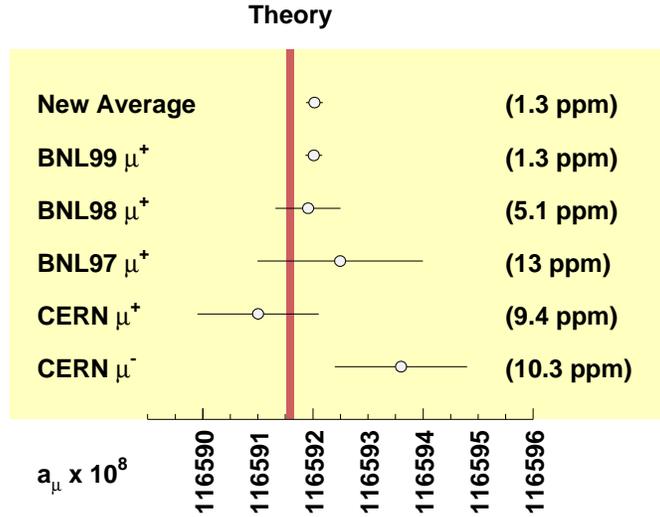}}
  \caption{Comparison of experimental and theoretical values of
           $a_\mu$.  \label{fig:confront}}
\end{figure}

In addition to $\omega_a$ and $B$, determination of $a_\mu$ would
require a value for the constant $e/m_\mu c$.  However, since the
magnetic field is measured by the (corrected) proton resonance
frequency $\omega_p$, the constant needed is $\lambda =
\mu_\mu/\mu_p$, so that $a_\mu$ is given by $a_\mu = R/(\lambda-R)$,
where $R = \omega_a/\omega_p$ and $\lambda$~= 3.18334539(10)
(0.03~ppm)~\cite{lambda}.  After the $\omega_p$ and $\omega_a$
analyses were finalized, separately and independently, the results
were combined to yield $a_{\mu^+}~= 11\,659\,202(14)(6)\times10^{-10}$
(1.3~ppm).  This in good agreement with all previous
measurements~\cite{carey,brown,cern3} (see fig.~\ref{fig:confront}).

Assuming CPT symmetry, the new world average is $a_{\mu}~=
11\,659\,203\,(15)~\times~10^{-10}$ (1.3~ppm), which deviates from the
Standard Model prediction $a_\mu^{SM}~= 11\,659\,159.6(6.7) \times
10^{-10}$ by 2.6 times the combined experimental and theoretical
error,
\begin{equation}
  a_{\mu}(\text{expt}) - a_{\mu}(\text{SM}) = 43(16)\times10^{-10}.
\end{equation}
Whether this tantalizing discrepancy persists or turns out to be a
statistical fluke will be found out in the near future.  In 2000,
approximately four times more positrons were recorded as in 1999.  In
2001, we completed our first measurement with negative muons, which
will provide a sensitive test of CPT violation.

\section*{Acknowledgments}

The author received financial support to attend this conference in the
form of a European Union ``Training and Mobility of Researchers
Programme'' grant.  The work described in this paper was supported in
part by the U.S.  Department of Energy, the U.S.  National Science
Foundation, the German Bundesminister f\"{u}r Bildung und Forschung,
the Russian Ministry of Science, and the US-Japan Agreement in High
Energy Physics.

\end{document}